# MANY-TO-MANY VOICE CONVERSION USING CYCLE-CONSISTENT VARIATIONAL AUTOENCODER WITH MULTIPLE DECODERS

*Keonnyeong Lee, In-Chul Yoo, and Dongsuk Yook*

Artificial Intelligence Laboratory, Department of Computer Science and Engineering
Korea University, Republic of Korea
{gnl0813, icyoo, yook}@ai.korea.ac.kr

## ABSTRACT

One of the obstacles in many-to-many voice conversion is the requirement of the parallel training data, which contain pairs of utterances with the same linguistic content spoken by different speakers. Since collecting such parallel data is a highly expensive task, many works attempted to use non-parallel training data for many-to-many voice conversion. One of such approaches is using the variational autoencoder (VAE). Though it can handle many-to-many voice conversion without the parallel training, the VAE based voice conversion methods suffer from low sound qualities of the converted speech. One of the major reasons is because the VAE learns only the self-reconstruction path. The conversion path is not trained at all. In this paper, we propose a cycle consistency loss for VAE to explicitly learn the conversion path. In addition, we propose to use multiple decoders to further improve the sound qualities of the conventional VAE based voice conversion methods. The effectiveness of the proposed method is validated using objective and the subjective evaluations.

## 1. INTRODUCTION

Voice conversion (VC) is a task of converting the speaker-related voice characteristics in an utterance while maintaining the linguistic information. Conventional VC methods require parallel speech data for the model training. The parallel speech data contain pairs of utterances that have the same linguistic contents spoken by different speakers. However, such parallel speech data are highly expensive that they restrict the use of VC in many applications. Therefore, many recent VC approaches attempted to use non-parallel training data. Early works using non-parallel training data adopt Gaussian mixture models (GMM) [1, 2, 3]. Recently, deep learning based VC approaches that have shown promising results use cycle-consistent adversarial networks (CycleGAN) [4, 5, 6, 7, 8], variational autoencoders (VAE) [9, 10, 11, 12], and VAE with generative adversarial networks (GAN) [13, 14].

In the CycleGAN [15] based VC approaches, the speech features of a source speaker are converted to match the characteristics of a target speaker using a GAN [16], and the converted speech features are again converted back through another GAN to match the original speech features from the source speaker. By using the cycle-consistency loss [17], the linguistic contents are forced to be retained in the converted speech. However, the CycleGAN can learn only one-to-one mapping between two speakers. To achieve complete mapping among $n$ speakers, $n(n-1)/2$ CycleGAN models must be trained separately, which increases the training time and the memory space prohibitively. Though the extensions of the CycleGAN for many-to-many VC have been proposed [6, 7, 8], they do not scale well as the number of speakers increases. For example, the number of speakers used in the experiments [6, 7, 8] were at most 4.

The VAE based VC approaches, on the other hand, can perform many-to-many VC for hundreds of speakers using non-parallel training data. A VAE [18] is composed of an encoder and a decoder. In the VC task, the encoder transforms the input speech features into the latent vectors containing the linguistic information of the input speech. Then, the latent vectors together with a target speaker identity vector, which is typically represented as a one-hot vector, are fed into the decoder to generate the converted speech features of the target speaker. Since the decoder is conditioned on a target speaker identity vector, it is sometimes called the conditional VAE.

Though the VAE models can be trained quickly, the sound qualities of the converted speech are usually low. To improve the sound quality, a VAE and Wasserstein generative adversarial network (WGAN) [19] hybrid called variational autoencoding Wasserstein generative adversarial networks (VAEWGAN) [13] was proposed. In this method, the decoder of the VAE is considered as the generator of a WGAN in order to train the decoder better. Though VAEWGAN based VC reduces some muffled sound, the qualities of the converted speech are still unsatisfactory.

One of the major drawbacks of the VAE based VC approaches is that the VAE models are not explicitly trained to convert the speech from a source speaker to a target speaker. Rather, they are trained to recover the same input speech from the source speaker using the latent vectors and the source speaker identity vector. In this paper, we propose to utilize a cycle consistency loss for the VAE to explicitly learn the mapping from a source speaker to a target speaker. To improve the sound quality further, we also propose a multi-decoder VAE which has a separate decoder for each target speaker. The cycle consistency loss and the multiple decoders can be incorporated into the VAEWGAN as well [20].

The rest of the paper is organized as follows. In Section 2, we describe the proposed methods in detail. Section 3 analyzes the experimental results, and Section 4 concludes the paper.

## 2. CYCLE-CONSISTENT VAE AND VAEWGAN

### 2.1. Variational Autoencoder

The loss function of the VAE is defined as follows:

$$\mathcal{L}_{\text{VAE}}(\phi,\theta;x,X) = \mathbb{D}_{\text{KL}}\left(q_\phi(z|x) \parallel p(z)\right) - \mathbb{E}_{z \sim q_\phi(z|x)}\left[\log(p_\theta(x|z,X))\right], \quad (1)$$

where $\mathbb{D}_{\text{KL}}$ is the Kullback-Leibler divergence, $q_\phi(z|x)$ is an encoding model with parameter $\phi$ that infers the linguistic information of input speech $x$, $p(z)$ is a prior distribution for latent vector $z$, and $p_\theta(x|z,X)$ is a decoding model with parameter $\theta$ that generates the reconstructed speech using $z$ and source speaker identity vector $X$.

To convert the speech from a source speaker to a target speaker, the source speaker identity vector $X$ is replaced with the target speaker identity vector $Y$. By minimizing equation (1), the VAE is trained to reconstruct the input speech from the latent vector $z$ and the source speaker identity vector $X$. Due to the absence of explicit model training for the conversion between the source speaker and the target speaker (i.e., only self-reconstruction training), the VAE based VC methods generally produce the converted speech with low sound quality.

### 2.2. Variational Autoencoder with Wasserstein Generative Adversarial Network

The VAEWGAN has been proposed to improve the sound quality of the VAE based VC method. In this approach, the decoder of the VAE is the generator of the WGAN. The loss function of the WGAN is defined as follows:

$$\mathcal{L}_{\text{WGAN}}(\theta,\psi;\phi,x,Y) = \mathbb{E}_y[D_\psi(y)] - \mathbb{E}_{z \sim q_\phi(z|x)}[D_\psi(G_\theta(z))], \quad (2)$$

where $G_\theta$ is the generator with parameter $\theta$, $D_\psi$ is the discriminator with parameter $\psi$, and $y$ is the speech from the target speaker represented by speaker identity vector $Y$. Since the decoder of the VAE is the generator of the WGAN, $G_\theta$ is $p_\theta$.

Now, the loss function of the VAEWGAN is defined as follows:

$$\mathcal{L}_{\text{VAEWGAN}}(\phi,\theta,\psi;x,Y) = \mathcal{L}_{\text{VAE}}(\phi,\theta;x,X) + \lambda_1 \mathcal{L}_{\text{WGAN}}(\theta,\psi;\phi,x,Y), \quad (3)$$

where $\lambda_1$ is the weight of the WGAN loss. Equation (3) is minimized for the VAE and the generator, and it is maximized for the discriminator. First, the VAE is trained in the same way as in Section 2.1. Second, the VAE and the WGAN are jointly trained such that the VAE gets an additional error signal from the discriminator of the WGAN.

Though the VAEWGAN produces somewhat higher sound quality than the VAE, it can handle only one-to-one voice conversion. In the next sections, we propose the extensions of the VAE and the VAEWGAN, called cycle-consistent VAE (CycleVAE) and cycle-consistent VAEWGAN (CycleVAEWGAN), respectively, which can improve the performance for many-to-many voice conversion by using multiple decoders and explicitly learning many-to-many mapping functions.

### 2.3. Cycle-Consistent Variational Autoencoder (CycleVAE)

In order to improve the sound quality of the VAE based VC, we propose to use a separate decoder for each speaker instead of a single decoder for all speakers. We also propose to use the cycle consistency loss for explicit conversion path training. The speaker identity vectors are not needed for the multi-decoder VAE since each speaker has an independent decoder. It can be expected that the sound quality can be improved since each decoder learns its corresponding speaker's voice characteristics by the additional conversion path training while the conventional VAE must cover multiple speakers with only a single decoder by self-reconstruction training.

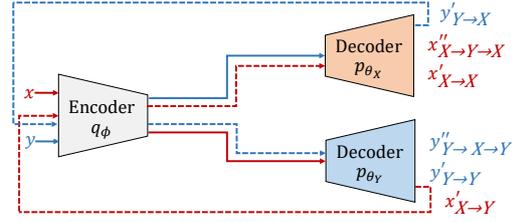

*Figure 1. CycleVAE.*

Fig. 1 shows the concept of the CycleVAE for two speakers. When the speech $x$ from speaker $X$ is fed into the network, it passes through the encoder and is compressed into the latent vector $z$. The reconstruction error is computed using the reconstructed speech $x'_{X \to X}$ by the speaker $X$ decoder model $p_{\theta_X}$. Up to this point, the loss function is similar to the vanilla VAE except that it does not require the speaker identity vectors, which is as follows:

$$\mathcal{L}'_{\text{VAE}}(\phi,\theta;x,X) = \mathbb{D}_{\text{KL}}\left(q_\phi(z|x) \parallel p(z)\right) - \mathbb{E}_{z \sim q_\phi(z|x)}\left[\log\left(p_{\theta_X}(x|z)\right)\right]. \quad (4)$$

The same input speech $x$ from speaker $X$ goes through the encoder and the speaker $Y$ decoder model $p_{\theta_Y}$ as well to generate the converted speech $x'_{X \to Y}$ which has the same linguistic contents as $x$ but in speaker $Y$'s voice. Then, the converted speech $x'_{X \to Y}$ goes through the encoder and $p_{\theta_X}$ to generate the converted back speech $x''_{X \to Y \to X}$ which should recover the original speech $x$. This cyclic conversion encourages the explicit training of voice conversion from $X$ to $Y$ and $Y$ to $X$. The cycle consistency loss of the multi-decoder VAE is defined as follows:

$$\mathcal{L}_{\text{cycle}}(\phi,\theta;x,X,Y) = \mathbb{D}_{\text{KL}}\left(q_\phi(z|x'_{X \to Y}) \parallel p(z)\right) - \mathbb{E}_{z \sim q_\phi(z|x'_{X \to Y})}\left[\log\left(p_{\theta_X}(x|z)\right)\right]. \quad (5)$$

Now, given the input speech $x$ from speaker $X$, the loss function of the CycleVAE for two speakers can be defined as follows:

$$\mathcal{L}_{\text{CycleVAE}}(\phi,\theta;x,X,Y) = \mathcal{L}'_{\text{VAE}}(\phi,\theta;x,X) + \lambda_2 \mathcal{L}_{\text{cycle}}(\phi,\theta;x,X,Y), \quad (6)$$

where $\lambda_2$ is the weight of the cycle consistency loss.

It can be easily extended for more than two speakers by summing over all pairs of the training speakers. The loss function of the CycleVAE for more than two speakers can be computed as follows for the input speech $x$ from speaker $X$:

$$\sum_Y \mathcal{L}_{\text{CycleVAE}}(\phi, \theta; x, X, Y) \, . \tag{7}$$

## 2.4. Cycle-Consistent Variational Autoencoder with Wasserstein Generative Adversarial Network (CycleVAEWGAN)

The CycleVAE can be extended to utilize the WGAN as in the VAEWGAN case. In the CycleVAEWGAN, the decoders of the CycleVAE are shared with the generators of the WGANs. Each decoder has its own WGAN. Fig. 2 shows the concept of the CycleVAEWGAN for two speakers. Since there are multiple WGANs, equation (2) is modified as follows:

$$\mathcal{L}'_{\text{WGAN}}(\theta, \psi; \phi, x, Y) = \mathbb{E}_y[D_{\psi_Y}(y)] - \mathbb{E}_{z \sim q_\phi(z|x)}\left[D_{\psi_Y}\left(G_{\theta_Y}(z)\right)\right], \tag{8}$$

where $G_{\theta_Y}$ is the generator with parameter $\theta_Y$ for speaker $Y$, $D_{\psi_Y}$ is the discriminator with parameter $\psi_Y$ for speaker $Y$, and $y$ is the speech from target speaker $Y$. Since the decoders of the CycleVAE are the generators of the WGANs, $G_{\theta_Y}$ is $p_{\theta_Y}$.

Now, the loss function of the CycleVAEWGAN is defined as follows:

$$\begin{aligned}\mathcal{L}_{\text{CycleVAEWGAN}}(\phi, \theta, \psi; x, X, Y) = \\ \mathcal{L}_{\text{CycleVAE}}(\phi, \theta; x, X, Y) + \\ \lambda_1 \mathcal{L}'_{\text{WGAN}}(\theta, \psi; \phi, x, X) + \\ \lambda_1 \mathcal{L}'_{\text{WGAN}}(\theta, \psi; \phi, x, Y) \, . \end{aligned} \tag{9}$$

Note that $\mathcal{L}'_{\text{WGAN}}$ is used twice in the equation, i.e., one for the self-reconstruction path and the other for the conversion path. Equation (9) is minimized for the CycleVAE and the generators, and is maximized for the discriminators. The first stage of the CycleVAEWGAN training is identical to the training procedure of the CycleVAE. In the second stage of the training, the CycleVAE and the WGANs are jointly optimized where the CycleVAE receives the additional error signals from the WGANs.

It also can be easily extended to more than two speakers by summing over all pairs of the training speakers. The loss function of the CycleVAEWGAN for more than two speakers can be computed as follows for the input speech $x$ from speaker $X$:

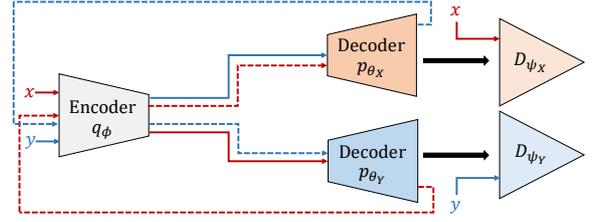

*Figure 2. CycleVAEWGAN.*

$$\sum_Y \mathcal{L}_{\text{CycleVAEWGAN}}(\phi, \theta, \psi; x, X, Y) \, . \tag{10}$$

The proposed CycleVAEWGAN is different from [12] in that it can utilize multiple decoders and WGANs.

## 3. EXPERIMENTS

For the experiments, 2 male speakers and 2 female speakers, namely SF1, SF2, TM1 and TM2, from VCC2018 dataset [21] were used. The numbers of the training and the testing utterances per speaker were 81 and 35, respectively. The speech were down-sampled to 22.05 kHz, and 36-dimensional Mel-frequency cepstral coefficients (MFCC), aperiodicities (AP), and fundamental frequency (F0) were extracted using the WORLD speech analyzer [22].

The encoders, the decoders, and the discriminators used the gated linear units (GLU) [23]. The batch normalization [24] was applied to each convolutional neural network (CNN) [25] layers. We built our models based on [11]. Fig. 3 shows the details of the encoder, decoder, and discriminator. We used the Adam optimizer [26] with a batch size of 8. $\lambda_1$ and $\lambda_2$ were set to 0 and 1, respectively, when training the CycleVAE. For the training of the CycleVAEWGAN, $\lambda_1$ and $\lambda_2$ were set to 1. All experiments were repeated 5 times starting with randomly initialized weights.

### 3.1. Objective Evaluations

One of drawbacks of the VAE based approaches is the over-smoothing of the generated data [27]. The global variance (GV) of MFCCs can be used to measure the degree of over-smoothing as the high GV values correlate with the sharpness of the spectra. We computed the GV for each of the MFCC indices. Fig. 4 shows the average GV over all evaluation utterances for the real speech and the converted speech by the conventional VAEWGAN and the proposed CycleVAE. The average GVs over all indices and all evaluation utterances were 0.247, 0.200, and 0.210 for the real speech and the converted speech by the VAEWGAN and CycleVAE, respectively.

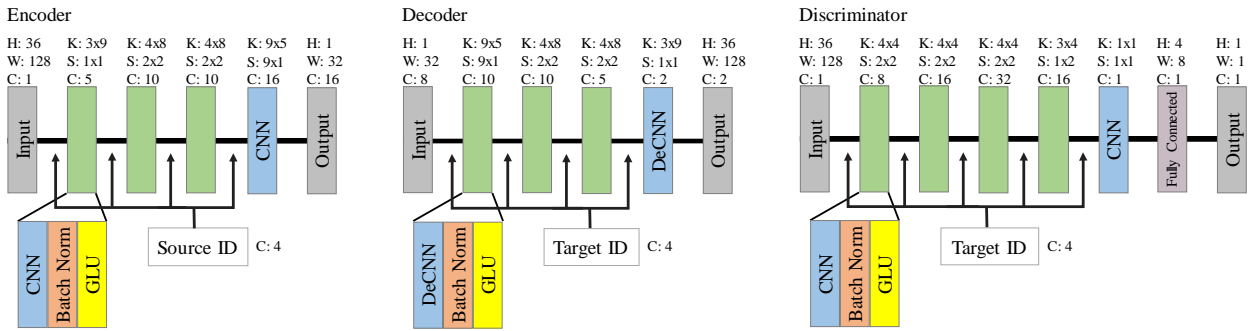

*Figure 3. The architectures of the encoder, decoder, and discriminator used in the experiments. The target speaker identity vectors (Target ID) are not used for the multi-decoder CycleVAE and CycleVAEWGAN.*

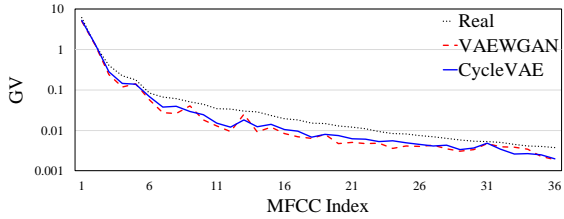

*Figure 4. Global variance of MFCCs for real speech utterances and the converted utterances by the VAE and the CycleVAE.*

For the case of the original and the converted speech utterances containing the same linguistic information, the difference between the MFCCs of the two speech utterances should be small. We used two metrics to measure this difference, i.e., the Mel-cepstral distortion (MCD) [27] and the modulation spectral distance (MSD) [28]. Tables 1 and 2 show MCD and MSD, respectively, for various VC methods. Firstly, by comparing the baseline VAE and VAEWGAN columns in the tables, we confirmed that the VAEWGAN outperforms the VAE [13]. Secondly, to measure the effectiveness of the cycle consistency loss alone, we built the CycleVAE and the CycleVAEWGAN that use a single common decoder (i.e., $p_{\theta_X}$ and $p_{\theta_Y}$ are shared in Fig. 1). The fourth and fifth columns of the tables shows these results. By comparing the second and the fourth columns (or the third and the fifth columns) of the tables, we confirmed the effectiveness of the cycle consistency loss [12]. Finally, the results of the proposed multi-decoder approaches with the cycle consistency loss are shown in the last two columns of the tables. It can be seen that the multi-decoder approaches improve the performances further. It is interesting to note that unlike the VAE or the CycleVAE having a single common decoder, adding the WGANs to the CycleVAE with multiple decoders does not improve the performance further. It is believed that because the multi-decoder cycle consistency loss is effective enough in learning the conversion path explicitly, the additional WGANs for conversion path learning may not be necessary.

### 3.2. Subjective Evaluations

We also conducted two subjective evaluations, i.e., naturalness test and similarity test. A set of 16 utterances was selected randomly such that four utterances were assigned to each pair of F to F, M to F, F to M, and M to M conversions, where F and M represent female and male, respectively. A total of 48 utterances (16 target speakers' utterances, 16 converted utterances by the conventional VAEWGAN, and 16 converted utterances by the proposed CycleVAE with multiple decoders) were played to 10 listeners participated in the subjective evaluations.

The mean opinion score (MOS) was used for the naturalness test. The listeners evaluated the naturalness of the speech in the scales of 1 (bad) to 5 (excellent) when the utterances were played in random order. Table 3 shows that the proposed CycleVAE based VC generally exhibits higher naturalness scores than the conventional VAEWGAN based VC.

In the similarity test, a target speaker's utterance was played first, then a pair of two converted utterances by the two methods were played in random order. The listeners were asked to select the more similar utterance to the target speaker's speech or 'fair' if they cannot tell the difference. Table 4 shows that the proposed CycleVAE based VC outperforms the conventional VAEWGAN based VC significantly.

*Table 1. MCD with standard deviation.*

|        | VAE | VAEWGAN | CycleVAE (single decoder) | CycleVAEWGAN (single decoder) | CycleVAE (multi-decoder) | CycleVAEWGAN (multi-decoder) |
|---|---|---|---|---|---|---|
| F to F | 7.31 ± 0.41 | 7.33 ± 0.38 | 7.20 ± 0.42 | 7.11 ± 0.43 | **6.97 ± 0.41** | 7.05 ± 0.41 |
| M to F | 7.75 ± 0.57 | 7.54 ± 0.52 | 7.45 ± 0.53 | 7.42 ± 0.52 | **7.23 ± 0.56** | 7.31 ± 0.52 |
| F to M | 7.32 ± 0.44 | 7.35 ± 0.40 | 7.17 ± 0.41 | 7.21 ± 0.39 | **7.03 ± 0.44** | 7.11 ± 0.43 |
| M to M | 7.40 ± 0.33 | 7.27 ± 0.31 | 7.17 ± 0.31 | 7.10 ± 0.30 | **7.00 ± 0.31** | 7.07 ± 0.31 |
| Average | 7.45 ± 0.44 | 7.37 ± 0.40 | 7.25 ± 0.42 | 7.21 ± 0.41 | **7.06** ± 0.43 | 7.13 ± 0.42 |

*Table 2. MSD with standard deviation.*

|        | VAE | VAEWGAN | CycleVAE (single decoder) | CycleVAEWGAN (single decoder) | CycleVAE (multi-decoder) | CycleVAEWGAN (multi-decoder) |
|---|---|---|---|---|---|---|
| F to F | 1.87 ± 0.16 | 1.85 ± 0.16 | 1.86 ± 0.16 | **1.84** ± 0.16 | 1.85 ± 0.15 | 1.85 ± 0.16 |
| M to F | 1.85 ± 0.14 | 1.83 ± 0.14 | 1.84 ± 0.14 | 1.83 ± 0.14 | **1.82** ± 0.13 | **1.82** ± 0.14 |
| F to M | 1.84 ± 0.17 | 1.83 ± 0.17 | 1.82 ± 0.16 | **1.80** ± 0.16 | 1.82 ± 0.17 | 1.83 ± 0.17 |
| M to M | 1.85 ± 0.16 | 1.84 ± 0.17 | 1.83 ± 0.17 | **1.82** ± 0.17 | **1.82** ± 0.16 | **1.82** ± 0.17 |
| Average | 1.86 ± 0.16 | 1.84 ± 0.16 | 1.84 ± 0.16 | **1.82** ± 0.16 | 1.83 ± 0.15 | 1.83 ± 0.16 |

*Table 3. Sound quality test (MOS and standard deviation).*

|        | VAEWGAN | CycleVAE | Target Voice |
|--------|---------|----------|--------------|
| F to F | 2.83 ± 0.86 | **3.08** ± 0.91 | 4.89 ± 0.39 |
| M to F | 2.15 ± 0.76 | **2.38** ± 0.91 | 4.89 ± 0.39 |
| F to M | 2.48 ± 0.92 | **2.65** ± 0.94 | 4.88 ± 0.40 |
| M to M | 2.38 ± 0.73 | **2.73** ± 0.87 | 4.88 ± 0.40 |
| Average | 2.46 ± 0.86 | **2.71** ± 0.94 | 4.88 ± 0.39 |

*Table 4. Similarity test (%).*

|        | VAEWGAN | Fair | CycleVAE |
|--------|---------|------|----------|
| F to F | 15.0 | 47.5 | 37.5 |
| M to F | 15.0 | 25.0 | 60.0 |
| F to M | 2.5  | 45.0 | 52.5 |
| M to M | 15.0 | 37.5 | 47.5 |
| Average | 12.0 | 39.0 | 49.0 |

## 4. CONCLUSION

In this paper, we proposed the new many-to-many voice conversion methods based on the VAE. The proposed methods use multiple decoders and explicitly learn the conversion path for many-to-many voice conversion. The effectiveness of the proposed methods was validated using the objective evaluations and the subjective evaluations.

The proposed methods can be further extended by utilizing multiple encoders, i.e., one encoder for each source speaker. Also, replacing the vocoder with powerful neural vocoders such as the WaveNet [29] or the WaveRNN [30] can be another future research direction.

## 5. ACKNOWLEDGEMENTS

This research was supported by the Basic Science Research Program through the National Research Foundation of Korea (NRF) funded by the Ministry of Science, ICT and Future Planning (NRF-2017R1E1A1A01078157). Also, it was partly supported by the MSIT (Ministry of Science and ICT) under the ITRC (Information Technology Research Center) support program (IITP-2018-0-01405) supervised by the IITP (Institute for Information & Communications Technology Planning & Evaluation), and IITP grant funded by the Korean government (MSIT) (No. 2018-0-00269).